\documentclass{article}
\usepackage{amssymb}
\usepackage{graphicx}
\usepackage{longtable}
\begin{document}

\def\tfbox#1{#1} 


\newdimen\younght    \younght=10pt
\newdimen\youngwd   \youngwd=10pt
\newdimen\youngthickness     \youngthickness=0.2pt

\def\beginYoung{
\begingroup
\def\vr{\vrule height0.8\younght width\youngthickness depth 0.2\younght}
\def\fbox##1{\vbox{\offinterlineskip
\hrule height\youngthickness
\hbox to \youngwd{\vr\hfill$##1$\hfill\vr}
\hrule height\youngthickness}}
\vbox\bgroup \offinterlineskip \tabskip=-\youngthickness \lineskip=-\youngthickness
\halign\bgroup &\fbox{##\unskip}\unskip  \crcr }

\def\End@Young{\egroup\egroup\endgroup}
\newenvironment{Young}{\beginYoung}{\End@Young}


\def\Tr#1{{{\rm Tr}\,#1}}
\def\tr#1{{{\rm tr}\,#1}}
\def\pp{{\phantom'}}
\def\Resp#1{{{\rm Res}^+_{#1}\,}}

\thispagestyle{empty}

\begin{flushright}
hep-th/0504230\\
ULB-TH 05-08
\end{flushright}

\bigskip

\begin{center}

{\bf\Large The structure of $E_{10}$ at higher $A_9$ levels -- a first
algorithmic approach}

\bigskip\bigskip

{\bf Thomas Fischbacher\\}

\smallbreak

{\em 
 Physique Th\'eorique et Math\'ematique and\\
 International Solvay Institutes, Universit\'e Libre de Bruxelles,\\
 Campus Plaine C.P. 231, B--1050 Bruxelles, Belgium\\[5mm]
}

{\small {\tt tf@cip.physik.uni-muenchen.de}}

\end{center}

\begin{abstract} 
\noindent The conjecture of a hidden $E_{10}$ symmetry of M-theory
is supported by the close connection between the dynamics of $D=11$
supergravity near a spacelike singularity and a truncation of an
one-dimensional $\sigma$-model with $E_{10}$ symmetry where all
representations beyond $SL(10)$ level $\ell=3$ are omitted. If this
conjecture is right, higher-level representations should especially
capture the dynamics of further M-theory degrees of
freedom. Unfortunately, the level by level determination of $E_{10}$
commutators which is necessary to extend the model to higher levels is
both an involved and toilsome task that requires computer aid. In this
work, some of the relevant problems are exposed and algorithmic
methods are developed which simplify key steps in the determination of
explicit $E_{10}$ commutators at higher levels.  As an application, we
compute the commutator of the level-two six-form with itself.
\end{abstract}

\section{Introduction}

\noindent Since Witten's discovery~\cite{Witten:1995ex}
that the strong coupling limit of the five consistent ten dimensional
superstring theories may be an {\em eleven}-dimensional theory whose
low energy regime is the unique known eleven-dimensional supergravity,
there has been a lot of activity trying to get a better understanding
of the nature of this strong coupling regime of string theory, as this
is is a highly promising candidate for the complete unified
description of the dynamics of all the fundamental forces. While a lot
could be derived about this theory in the last decade, our picture is
still quite incomplete, and this is mirrored by the preliminary
incomplete name {\em M-theory}.

As the study of symmetries -- possibly hidden and showing up only in
certain limits of a model -- has been extremely fruitful for the
advancement of our understanding of fundamental physics, during the
last century, often leading to a major breakthrough, any indication of
a not yet well understood symmetry of M-theory should be considered as
a potential important clue, and subjected to close
investigation. While the existence of hidden exceptional symmetries of
eleven-dimensional supergravity, especially the hidden global
$E_{11-d}$ symmetry that occurs in Kaluza-Klein reduction on a
$d\le8$-dimensional torus, is well known (see
e.g. \cite{Cremmer:1997ct} for an introduction), and speculations
about a hidden global {\em infinite-dimensional} Kac-Moody symmetry
also have been around for a long time~\cite{julia}, the conjecture of
a fundamental $E_{10}$ symmetry of M-theory recently gained interest
by the discovery that for the bosonic sector of eleven-dimensional
supergravity in a setting close to a spacelike `big bang' singularity
(where spatially separated points become causally disconnected), the
degrees of freedom of the metric show chaotic oscillations of BKL
type~\cite{Belinsky:1982pk} that can be described by relativistic
billiards~\cite{Damour:2000wm,Damour:2000hv,Damour:2001sa} taking place
in a region of the shape of the fundamental Weyl chamber of
$E_{10}$. Intriguingly, a one-dimensional
$\sigma$-model~\cite{Damour:2002cu} constructed from a truncation of
$E_{10}$ to the first three levels of a $SL(10)$
decomposition~\cite{Nicolai:2003fw} is able to capture and reproduce
the dynamics of a correspondingly truncated expansion in terms of
gradients of $D=11$ supergravity in the previously described
limit. While this already is a highly nontrivial result, one may still
be sceptical whether this correspondence may be extended to higher
levels. At least, $E_{10}$ seems to contain all the fields necessary
to represent higher gradients of the $3$-form and $6$-form potential
as well as the `dual' graviton. Besides those, there are many other
representations that are conjectured to be linked to other dynamical
(potentially even yet unknown) degrees of freedom of M-theory. First
promising indications that such a description of M-theory via an
$E_{10}$ $\sigma$-model may indeed work are given by the highly
nontrivial observation that it may be reconciled (in a truncation to
low levels) with both (massive) $IIA$ and $IIB$
supergravity~\cite{Kleinschmidt:2004dy,Kleinschmidt:2004rg}, as well
as the possible correspondence of $SL(10)$ singlets at quite high
levels to $R^4$, $R^7$,\dots corrections of M-theory~\cite{Damour:2005zb}.

Clearly, it is essential to perform an extension of the construction
beyond level three. As we will see, this is bound to require some
quite involved algorithms, and hence this line of research may well
benefit from scientific exchange with computer scientists specialized
in symbolic algebra. Therefore, the present article is somewhat more
verbose than one targeted exclusively at a string theory
audience. While it is not expected of readers with a specialization in
symbolic algebra to comprehend all of the physics upon first reading,
at least the relation between the relevant concepts should become
clear. While we will be concerned exclusively with the determination
of commutators for the Borel subalgebra of $E_{10}$ in this work,
details on the construction of the model up to level three can be
found in~\cite{Damour:2004zy}.

\section{Conventions}

\noindent In the following, we will make extensive use of irreducible
representations of the $A_n=SL(n+1)$ Lie algebra that are obtained by
applying mixed-symmetry Young projectors on tensor powers of the
fundamental vector representation. We use the usual convention of
first symmetrizing over rows, then symmetrizing over columns of a
Young tableau, i.e.

\begin{equation}
P\Bigl(\kern-1em\raise-2ex\hbox{\begin{Young}a&b\cr c\cr\end{Young}}\Bigr)A^{abc}=\mbox{const.}\cdot\left((A^{abc}+A^{bac})-(a\leftrightarrow c)\right)
\end{equation}

It is furthermore convenient to lay out all the indices of a tensor
carrying a single irreducible representation in a form that directly
mirrors the shape of the corresponding symmetrizer's tableau. However,
as we will mostly have to deal with short symmetrizations and long
anti-symmetrizations in the analysis of $E_{10}$, we will rotate our
tableaux by a counterclockwise quarter-turn. Thus, {\em from now on},
we write e.g.
\begin{equation}
\begin{array}{lll}
{E_3^{\small\begin{Young}a\cr b&c&d&e&f&g&h&i\cr\end{Young}}}&\mbox{instead of}&E_3^{\small\begin{Young}b&a\cr c\cr d\cr e\cr f\cr g\cr h\cr i\cr\end{Young}}
\end{array}
\end{equation}
for the `dual graviton' generator at level three in the $A_9$
decomposition of $E_{10}$. {\em Despite this notation, which is
rotated solely for obvious typographic reasons, we will still speak of
the {\em symmetrization of rows} and {\em anti-symmetrization of
columns}!}

We will also extensively apply symmetrization and anti-symmetrization
operations to groups of indices. While this is usually denoted with
brackets or parentheses around groups of indices, such as
in\footnote{As usual, we define anti-symmetrizers as projectors, e.g.\hfill\break
$\delta^{abc}_{def}=\frac{1}{3!}\left(\delta^a_d\delta^b_e\delta^c_f\pm\rm{permutations}\right)$.
}
\begin{equation}
X_{a[bcd]e}:=X_{ab'c'd'e}\delta^{b'c'd'}_{b\pp c\pp d\pp}
\end{equation}
this ad-hoc notation has two important drawbacks: 
\begin{itemize}
\item It is highly unpractical for more complicated situations
 with interlocking symmetrizations.

\item Positional information is used to convey information
 about symmetrizations.
\end{itemize}

In particular, there is no reasonable way to translate an expression like
\begin{equation}
X_{a'b'c'd'e'f'g'}\delta^{a'd'g'}_{a\pp d\pp g\pp}\delta^{b'f'}_{b\pp f\pp}\delta^{c'e'}_{c\pp e\pp}
\end{equation}
to full bracketed-indices symmetrization notation. Such quantities
are, however, bound to appear in the analysis of higher levels of
hyperbolic Kac-Moody symmetries.

The second problem mentioned has an important conceptual aspect: there
are two different ways to apply an (anti-)symmetrization to a given
mixed-symmetry tensor. One has to discern between using the
antisymmetry in indices of the same column to exchange boxes, i.e.
\begin{equation}
X^{\begin{Young}*\cr*&*_1&*_2&*\cr\end{Young}}
\rightarrow
-X^{\begin{Young}*\cr*&*_2&*_1&*\cr\end{Young}}
\end{equation}
and applying a symmetrization to index names, as in:
\begin{equation}
X^{\begin{Young}a\cr b&c&d&e\cr\end{Young}}
\begin{array}{c}
{\scriptstyle \cdot\delta_{a\pp b\pp}^{a'b'}}\\
\longrightarrow\\
{\scriptstyle a',b'\rightarrow a,b}
\end{array}
\frac{1}{2}
\left(
X^{\begin{Young}a\cr b&c&d&e\cr\end{Young}}
-X^{\begin{Young}b\cr a&c&d&e\cr\end{Young}}
\right)
\end{equation}
which would correspond to `placing brackets around $[ab]$'.

One can make the idea rigorous that the first operation acts on the
right, while the second one acts on the left of the operation of {\em
naming} tensor indices, hence we will denote the first kind of
symmetrization as a {\em box shifting} symmetrization, and the second
one as an {\em index relabeling} symmetrization. Note that box
shifting symmetrizations are conceptually tied to individual columns
of a single (mixed-symmetry) tensor, while index relabeling symmetries
are more naturally associated to entire sums, not individual summands,
as they amount to introducing an extra $\delta^{\ldots}_{\ldots}$
factor. A large part of the algebraic complexity of the higher levels
is related to the fundamental difference between these two concepts.

We clearly need a new notation for index relabeling symmetrizations
that overcomes the forementioned problems. We will use the convention
to denote symmetrization in indices $i,j,k,\ldots,n$ by appending a
formal (pseudo-)factor $\sigma(ijk\ldots n)$ to the right of the term,
and likewise anti-symmetrization in these indices by a formal factor
$\alpha(ijk\ldots n)$. This definition implicitly contains the rule to
apply such factors one by one from the left to the right to a term,
replacing $\alpha(ijk\ldots n)$ by $\delta^{i\pp j\pp k\pp \ldots
n\pp}_{i'j'k'\ldots n'}$ , where $i',j',k',\ldots,n'$ are fresh
indices, and substituting free indices $i\mapsto i'$, $j\mapsto j'$,
etc. in all factors to the left of it. If it helps to make a point, we
will occasionally include trivial one-index symmetrization factors
like~$\alpha(g)$ in our formulae.

To give an example, applying a specific Young projector may now be
written as:
\begin{equation}
P^{\small\begin{Young}a\cr b&c&d\cr\end{Young}} \langle{\mbox term}\rangle = 
\frac{8}{5}\langle term\rangle\cdot\sigma(ab)\cdot\alpha(bcd).
\end{equation}
The numerical factor $c_P$, $8/5$ in our example, that makes a
sequence of a tableau symmetrization followed by a tableau
anti-symmetrization an idempotent projector, is given by
\begin{equation}
c_P=\frac{\left(\prod_{r\in\rm rows} (\#r)!\right)\cdot\left(\prod_{c\in\rm columns} (\#c)!\right)}{\prod_{h\in\rm hooks}\#h}.
\end{equation}
(For every box in the tableau, there is exactly one hook, which
consists of this box and all that lie either in the same row and to
the right of it, or in the same column and further down.)

Note that with this notation, it is very easy to make proper use of
both box shifting and index relabeling symmetries. For example, we have:
\begin{equation}
\begin{array}{l}
\phantom{=+}X^{\begin{Young}a\cr b&c&d&e\cr\end{Young}}\cdot\alpha(a,b,c)\\
{=-}X^{\begin{Young}a\cr b&d&c&e\cr\end{Young}}\cdot\alpha(a,b,c)\qquad\mbox{(box shifting)}\\
{=+}X^{\begin{Young}c\cr b&d&a&e\cr\end{Young}}\cdot\alpha(a,b,c)\qquad\mbox{(index relabeling)}\\
{=+}X^{\begin{Young}c\cr a&b&d&e\cr\end{Young}}\cdot\alpha(a,b,c)\qquad\mbox{($2\times$ box shifting)}\\
\\
=\ldots
\end{array}
\end{equation}

This raises the question how to define normal forms for such
expressions. This is an involved issue that has many technical and
algorithmic aspects and hence is discussed in detail in the appendix
in section~\ref{normalforms}. In brief, the idea behind the scheme
used in this work is to treat all extra (non-overlapping) index
symmetrizations on a tensor as being ordered lexicographically, then
using them one after another to put as many lexically small tensor
indices into columns to the far right as possible. This means in
particular (as is shown with an example in the appendix) that
submitting a mixed-symmetry tensor in normal form to its associated
Young symmetrization will in general produce a sum of {\em more than
one} tensor in normal form. This is mostly due to our ignorance of
more complicated relations between mixed symmetry representations with
permuted indices, which are most easily expressed in terms of Garnir
symmetries~\cite{garnir} -- these will be explained in some more
detail at the end of section~\ref{l3mod}.

Defining a normal form that takes into account all three types of
column antisymmetrization, index relabeling, and Garnir symmetries,
and can be computed with sufficient efficiency for an application to
$E_{10}$ is not as inextricable as it may seem at first, but will not
be discussed here, even if using such a normal form that differs from
the one used here will eventually turn out to be inevitable when going
to high levels. Note that if one would not have to deal with index
relabeling symmetries, it would be possible to define normal forms of
filled tableaux just such that in a tensor in normal form, all indices
appear in lexically ascending order in every column as well as in
every row. Any tensor can be transformed into a linear combination of
tensors of that form by repeated application of Garnir as
well as column antisymmetries. It is easy to see that this
prescription has to be refined further to get unique normal forms in
the presence of index relabeling symmetries, as is demonstrated by the
following example:
\begin{equation}
X^{\begin{Young}b\cr a&c\cr\end{Young}}\cdot\sigma(bc)
=X^{\begin{Young}c\cr a&b\cr\end{Young}}\cdot\sigma(bc).
\end{equation}

One can imagine that the situation will get far more involved in
situations where e.g. index relabeling symmetries touch indices that
are not located at the boundary of the tableau.

A further issue concerns the use of epsilon tensors: in Young tableau
index notation, an element of the $E_{10}$ algebra which appears in
the $n$-fold iterated commutator
$[[\ldots\underbrace{[E_1^{abc},E_1^{def}],E_1^{ghi}],E_1^{jkl}]\ldots]}_{n}$
will carry $3n$ indices. Typically, some of those will form columns of
maximal length (here $10$) that could be extracted as extra factors
$\epsilon^{abc\ldots j}$. For the sake of a homogeneous presentation,
we will not do so, and instead keep all columns of maximal length
explicit.

\section{The $E_{10}$ Borel subalgebra from the $A_{9}$ perspective}

\noindent We start this section by reviewing a few ideas and facts
from Cartan-Dynkin theory and its application to $E_{10}$.
The roots of the $E_{10}$ algebra lie on the unique even unimodular
Lorentzian lattice $II_{9,1}$. Taking the ten-dimensional scalar
product to be
\begin{equation}
(x,x)=-x_{10}^2+\sum_{j=1}^{9}x_j^2,
\end{equation}
this lattice is spawned by these ten vectors of squared length two
(cf.~\cite{Barwald:1996wb})
\begin{equation}
\begin{array}{lccrrrrrrrrrrc}
r_1&=&(&\frac{1}{2}&\frac{1}{2}&\frac{1}{2}&\frac{1}{2}&\frac{1}{2}&\frac{1}{2}&\frac{1}{2}&\frac{1}{2}&\frac{1}{2}&\frac{1}{2}&)\\
r_2&=&(&-1&-1&0&0&0&0&0&0&0&0&)\\
r_3&=&(&0&1&-1&0&0&0&0&0&0&0&)\\
r_4&=&(&0&0&1&-1&0&0&0&0&0&0&)\\
r_5&=&(&0&0&0&1&-1&0&0&0&0&0&)\\
r_6&=&(&0&0&0&0&1&-1&0&0&0&0&)\\
r_7&=&(&0&0&0&0&0&1&-1&0&0&0&)\\
r_8&=&(&0&0&0&0&0&0&1&-1&0&0&)\\
r_9&=&(&0&0&0&0&0&0&0&1&-1&0&)\\
r_0&=&(&1&-1&0&0&0&0&0&0&0&0&)\\
\end{array}
\end{equation}

\begin{figure}
\begin{center}
\includegraphics[width=8cm]{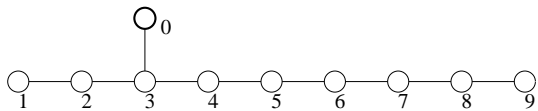}
\caption{\label{e10dynkin}The Dynkin diagram of $A_{9}$ extended to $E_{10}$}
\end{center}
\end{figure}

The Dynkin diagram encoding the geometrical structure of this lattice
is given in figure~\ref{e10dynkin}.

\subsection{Kac-Moody Algebras from Cartan matrices}

\noindent One key property of Lie algebras is the geometry
of their root lattice. Choosing a hyperplane through the origin that
does not contain any further lattice point in order to separate
lattice vectors into a `positive' and `negative' half and denoting
positive lattice vectors that cannot be written as the sum of other
positive lattice vectors as {\em simple roots}, this information can
be concisely encoded in the Cartan matrix $A_{ij}$, which contains the
scaled (not necessarily euclidean) scalar products of the simple
roots:
\begin{equation}
A_{ij}=2\frac{(\alpha_i,\alpha_j)}{(\alpha_j,\alpha_j)}
\end{equation}

The off-diagonal entries of a Cartan matrix are nonpositive integers,
due to the indecomposability of the simple roots. Vice versa, it is
possible to construct the Lie algebra $\mathfrak{g}_A$ from a
generalized Cartan matrix: one first defines the Chevalley generators
$e_i,f_i,h_i$, with commutators satisfying
\begin{equation}
\begin{array}{lclclcl}
{}[h_i,e_j]&=&A_{ij}e_j&\quad&[h_i,f_j]&=&-A_{ij}f_j\\
{}[e_i,f_j]&=&\delta_{ij}h_i&\quad&[h_i,h_j]&=&0.
\end{array}
\end{equation}

The algebra $\mathfrak{g}_A$ is then obtained first forming the free Lie
algebra, which is spawned by all the commutators that can be built
recursively from these generators, modulo the Jacobi identity
\begin{equation}
[[X,Y],Z]+[[Y,Z],X]+[[Z,X],Y]=0.
\end{equation}
This free Lie algebra then has to be divided by the Serre relations:
\begin{equation}
\begin{array}{l}
{}\underbrace{[e_i,[e_i,\ldots,[e_i}_{1-A_{ij}},e_j]\ldots]]=0\\
{}\underbrace{[f_i,[f_i,\ldots,[f_i}_{1-A_{ij}},f_j]\ldots]]=0.
\end{array}
\end{equation}

For finite-dimensional Lie algebras, all roots have squared length
$\Lambda^2=2$. In the hyperbolic case -- where we will take $E_{10}$
as our primary example -- the metric on the root space is indefinite,
and while there is just one generator associated to every root with
$\Lambda^2=2$ (as in the finite-dimensional case), the multiplicity of
roots inside the `light cone', with $\Lambda^2\le0$, increases quite
dramatically with increasing $-\Lambda^2$. While the Jacobi and Serre
relations determine the structure of the Lie algebra, it is a
difficult and basically still unsolved problem to effectively identify
those multiple commutators of Chevalley generators $E_i$ that vanish
due to the Serre relations, which may be revealed only after multiple
applications of the Jacobi identity.

While one may construct a Lie algebra basis as well as determine the
structure constants directly from the simple roots, it is frequently
more convenient and more useful to identify a simple subalgebra that
has a simple geometrical description by omitting one (or more than
one) simple root, and slice the algebra into irreducible
representations of this sub-algebra, graded by the number of
occurrences of the operator $E_j$ or $F_j$ corresponding to the
omitted root $r_j$. This leads to the presentation of commutators in
the language of tensors of the sub-algebra. An example of this
technique, the decomposition of $E_8$ in terms of its $SL(8)=A_7$
sub-algebra, can be found in appendix~B of~\cite{Koepsell:2000xg}.

In this work, we limit our analysis to $E_{10}$ Borel subalgebra
commutation relations that are schematically of the form
\begin{equation}
E_{\ell=p+q}=\sum [E_{\ell=p},E_{\ell=q}].
\end{equation}
Most of the technical complication comes from the demand to work not
with explicit roots, but abstract tensor equations, and the
combinatorical infeasibility to use naive (anti-)symmetrization by
generating all permutations if more than eight indices have to be
(anti)-symmetrized. Hence, the question arises how to organize
calculations in such a way that maximal use can be made of overlaps
with pre-existing symmetries.

\subsection{The $A_9$ decomposition of $E_{10}$ at level $\ell\le3$}

\noindent The determination of commutators for the $E_{10}$ algebra
not only gets more and more voluminous with increasing $A_9$ level,
one also has to use an increasing number of different algebraic tricks
with increasing level. Also, some steps performed for the first few
levels of the construction can be interpreted in multiple -- sometimes
unexpected -- ways, and often, only one interpretation can be suitably
generalized to be of use for higher levels as well.

Hence, it makes sense to start the discussion of higher levels with a
brief review of the derivation of commutators up to level~$\ell=3$,
where we mark those steps in the construction that require closer
inspection, resp. modification, when going to higher levels with
$(*_n)$. A more detailed discussion of the nature of these steps will
then be given in section~\ref{l3mod}.

One important tool in the derivation of commutation relations is the
representation table in~\cite{Nicolai:2003fw}, whose first few entries
we quote for convenience:

\begin{longtable}{|r|r|r|r|r|r|r|}
\caption{\label{e10table}$A_9$ representations in $E_{10}$ up to level $\ell=6$}\\
\hline
\multicolumn{1}{|c|}{$\ell$}&\multicolumn{1}{c|}{$p$}&\multicolumn{1}{c|}{$m$}&\multicolumn{1}{c|}{$\Lambda^2$}&\multicolumn{1}{c|}{dim ${\mathcal{R}}(\Lambda)$}&\multicolumn{1}{c|}{${\rm mult}(\Lambda)$}&\multicolumn{1}{c|}{$\mu$}\\
\hline
\hline
$1$&(\hbox to 0.5em{\hfill 0}\hbox to 0.5em{\hfill 0}\hbox to 0.5em{\hfill 1}\hbox to 0.5em{\hfill 0}\hbox to 0.5em{\hfill 0}\hbox to 0.5em{\hfill 0}\hbox to 0.5em{\hfill 0}\hbox to 0.5em{\hfill 0}\hbox to 0.5em{\hfill 0})&\kern-0.7em\hbox to 1.5em{\hfill 0}\hbox to 1.5em{\hfill 0}\hbox to 1.5em{\hfill 0}\hbox to 1.5em{\hfill 0}\hbox to 1.5em{\hfill 0}\hbox to 1.5em{\hfill 0}\hbox to 1.5em{\hfill 0}\hbox to 1.5em{\hfill 0}\hbox to 1.5em{\hfill 0}&2&120&1&1\\
\hline
$2$&(\hbox to 0.5em{\hfill 0}\hbox to 0.5em{\hfill 0}\hbox to 0.5em{\hfill 0}\hbox to 0.5em{\hfill 0}\hbox to 0.5em{\hfill 0}\hbox to 0.5em{\hfill 1}\hbox to 0.5em{\hfill 0}\hbox to 0.5em{\hfill 0}\hbox to 0.5em{\hfill 0})&\kern-0.7em\hbox to 1.5em{\hfill 1}\hbox to 1.5em{\hfill 2}\hbox to 1.5em{\hfill 3}\hbox to 1.5em{\hfill 2}\hbox to 1.5em{\hfill 1}\hbox to 1.5em{\hfill 0}\hbox to 1.5em{\hfill 0}\hbox to 1.5em{\hfill 0}\hbox to 1.5em{\hfill 0}&2&210&1&1\\
\hline
$3$&(\hbox to 0.5em{\hfill 1}\hbox to 0.5em{\hfill 0}\hbox to 0.5em{\hfill 0}\hbox to 0.5em{\hfill 0}\hbox to 0.5em{\hfill 0}\hbox to 0.5em{\hfill 0}\hbox to 0.5em{\hfill 0}\hbox to 0.5em{\hfill 1}\hbox to 0.5em{\hfill 0})&\kern-0.7em\hbox to 1.5em{\hfill 1}\hbox to 1.5em{\hfill 3}\hbox to 1.5em{\hfill 5}\hbox to 1.5em{\hfill 4}\hbox to 1.5em{\hfill 3}\hbox to 1.5em{\hfill 2}\hbox to 1.5em{\hfill 1}\hbox to 1.5em{\hfill 0}\hbox to 1.5em{\hfill 0}&2&440&1&1\\
&(\hbox to 0.5em{\hfill 0}\hbox to 0.5em{\hfill 0}\hbox to 0.5em{\hfill 0}\hbox to 0.5em{\hfill 0}\hbox to 0.5em{\hfill 0}\hbox to 0.5em{\hfill 0}\hbox to 0.5em{\hfill 0}\hbox to 0.5em{\hfill 0}\hbox to 0.5em{\hfill 1})&\kern-0.7em\hbox to 1.5em{\hfill 2}\hbox to 1.5em{\hfill 4}\hbox to 1.5em{\hfill 6}\hbox to 1.5em{\hfill 5}\hbox to 1.5em{\hfill 4}\hbox to 1.5em{\hfill 3}\hbox to 1.5em{\hfill 2}\hbox to 1.5em{\hfill 1}\hbox to 1.5em{\hfill 0}&0&10&8&0\\
\hline
$4$&(\hbox to 0.5em{\hfill 0}\hbox to 0.5em{\hfill 0}\hbox to 0.5em{\hfill 1}\hbox to 0.5em{\hfill 0}\hbox to 0.5em{\hfill 0}\hbox to 0.5em{\hfill 0}\hbox to 0.5em{\hfill 0}\hbox to 0.5em{\hfill 0}\hbox to 0.5em{\hfill 1})&\kern-0.7em\hbox to 1.5em{\hfill 2}\hbox to 1.5em{\hfill 4}\hbox to 1.5em{\hfill 6}\hbox to 1.5em{\hfill 5}\hbox to 1.5em{\hfill 4}\hbox to 1.5em{\hfill 3}\hbox to 1.5em{\hfill 2}\hbox to 1.5em{\hfill 1}\hbox to 1.5em{\hfill 0}&2&1155&1&1\\
&(\hbox to 0.5em{\hfill 2}\hbox to 0.5em{\hfill 0}\hbox to 0.5em{\hfill 0}\hbox to 0.5em{\hfill 0}\hbox to 0.5em{\hfill 0}\hbox to 0.5em{\hfill 0}\hbox to 0.5em{\hfill 0}\hbox to 0.5em{\hfill 0}\hbox to 0.5em{\hfill 0})&\kern-0.7em\hbox to 1.5em{\hfill 1}\hbox to 1.5em{\hfill 4}\hbox to 1.5em{\hfill 7}\hbox to 1.5em{\hfill 6}\hbox to 1.5em{\hfill 5}\hbox to 1.5em{\hfill 4}\hbox to 1.5em{\hfill 3}\hbox to 1.5em{\hfill 2}\hbox to 1.5em{\hfill 1}&2&55&1&1\\
&(\hbox to 0.5em{\hfill 0}\hbox to 0.5em{\hfill 1}\hbox to 0.5em{\hfill 0}\hbox to 0.5em{\hfill 0}\hbox to 0.5em{\hfill 0}\hbox to 0.5em{\hfill 0}\hbox to 0.5em{\hfill 0}\hbox to 0.5em{\hfill 0}\hbox to 0.5em{\hfill 0})&\kern-0.7em\hbox to 1.5em{\hfill 2}\hbox to 1.5em{\hfill 4}\hbox to 1.5em{\hfill 7}\hbox to 1.5em{\hfill 6}\hbox to 1.5em{\hfill 5}\hbox to 1.5em{\hfill 4}\hbox to 1.5em{\hfill 3}\hbox to 1.5em{\hfill 2}\hbox to 1.5em{\hfill 1}&0&45&8&0\\
\hline
$5$&(\hbox to 0.5em{\hfill 0}\hbox to 0.5em{\hfill 0}\hbox to 0.5em{\hfill 0}\hbox to 0.5em{\hfill 0}\hbox to 0.5em{\hfill 0}\hbox to 0.5em{\hfill 1}\hbox to 0.5em{\hfill 0}\hbox to 0.5em{\hfill 0}\hbox to 0.5em{\hfill 1})&\kern-0.7em\hbox to 1.5em{\hfill 3}\hbox to 1.5em{\hfill 6}\hbox to 1.5em{\hfill 9}\hbox to 1.5em{\hfill 7}\hbox to 1.5em{\hfill 5}\hbox to 1.5em{\hfill 3}\hbox to 1.5em{\hfill 2}\hbox to 1.5em{\hfill 1}\hbox to 1.5em{\hfill 0}&2&1848&1&1\\
&(\hbox to 0.5em{\hfill 1}\hbox to 0.5em{\hfill 0}\hbox to 0.5em{\hfill 0}\hbox to 0.5em{\hfill 1}\hbox to 0.5em{\hfill 0}\hbox to 0.5em{\hfill 0}\hbox to 0.5em{\hfill 0}\hbox to 0.5em{\hfill 0}\hbox to 0.5em{\hfill 0})&\kern-0.7em\hbox to 1.5em{\hfill 2}\hbox to 1.5em{\hfill 5}\hbox to 1.5em{\hfill 8}\hbox to 1.5em{\hfill 6}\hbox to 1.5em{\hfill 5}\hbox to 1.5em{\hfill 4}\hbox to 1.5em{\hfill 3}\hbox to 1.5em{\hfill 2}\hbox to 1.5em{\hfill 1}&2&1848&1&1\\
&(\hbox to 0.5em{\hfill 0}\hbox to 0.5em{\hfill 0}\hbox to 0.5em{\hfill 0}\hbox to 0.5em{\hfill 0}\hbox to 0.5em{\hfill 1}\hbox to 0.5em{\hfill 0}\hbox to 0.5em{\hfill 0}\hbox to 0.5em{\hfill 0}\hbox to 0.5em{\hfill 0})&\kern-0.7em\hbox to 1.5em{\hfill 3}\hbox to 1.5em{\hfill 6}\hbox to 1.5em{\hfill 9}\hbox to 1.5em{\hfill 7}\hbox to 1.5em{\hfill 5}\hbox to 1.5em{\hfill 4}\hbox to 1.5em{\hfill 3}\hbox to 1.5em{\hfill 2}\hbox to 1.5em{\hfill 1}&0&252&8&0\\
\hline
$6$&(\hbox to 0.5em{\hfill 1}\hbox to 0.5em{\hfill 0}\hbox to 0.5em{\hfill 0}\hbox to 0.5em{\hfill 0}\hbox to 0.5em{\hfill 0}\hbox to 0.5em{\hfill 0}\hbox to 0.5em{\hfill 0}\hbox to 0.5em{\hfill 1}\hbox to 0.5em{\hfill 1})&\kern-0.7em\hbox to 1.5em{\hfill 3}\hbox to 1.5em{\hfill 7}\hbox to 1.5em{\hfill 11}\hbox to 1.5em{\hfill 9}\hbox to 1.5em{\hfill 7}\hbox to 1.5em{\hfill 5}\hbox to 1.5em{\hfill 3}\hbox to 1.5em{\hfill 1}\hbox to 1.5em{\hfill 0}&2&3200&1&1\\
&(\hbox to 0.5em{\hfill 0}\hbox to 0.5em{\hfill 0}\hbox to 0.5em{\hfill 0}\hbox to 0.5em{\hfill 0}\hbox to 0.5em{\hfill 0}\hbox to 0.5em{\hfill 0}\hbox to 0.5em{\hfill 0}\hbox to 0.5em{\hfill 0}\hbox to 0.5em{\hfill 2})&\kern-0.7em\hbox to 1.5em{\hfill 4}\hbox to 1.5em{\hfill 8}\hbox to 1.5em{\hfill 12}\hbox to 1.5em{\hfill 10}\hbox to 1.5em{\hfill 8}\hbox to 1.5em{\hfill 6}\hbox to 1.5em{\hfill 4}\hbox to 1.5em{\hfill 2}\hbox to 1.5em{\hfill 0}&0&55&8&0\\
&(\hbox to 0.5em{\hfill 0}\hbox to 0.5em{\hfill 1}\hbox to 0.5em{\hfill 0}\hbox to 0.5em{\hfill 0}\hbox to 0.5em{\hfill 0}\hbox to 0.5em{\hfill 1}\hbox to 0.5em{\hfill 0}\hbox to 0.5em{\hfill 0}\hbox to 0.5em{\hfill 0})&\kern-0.7em\hbox to 1.5em{\hfill 3}\hbox to 1.5em{\hfill 6}\hbox to 1.5em{\hfill 10}\hbox to 1.5em{\hfill 8}\hbox to 1.5em{\hfill 6}\hbox to 1.5em{\hfill 4}\hbox to 1.5em{\hfill 3}\hbox to 1.5em{\hfill 2}\hbox to 1.5em{\hfill 1}&2&8250&1&1\\
&(\hbox to 0.5em{\hfill 1}\hbox to 0.5em{\hfill 0}\hbox to 0.5em{\hfill 0}\hbox to 0.5em{\hfill 0}\hbox to 0.5em{\hfill 0}\hbox to 0.5em{\hfill 0}\hbox to 0.5em{\hfill 1}\hbox to 0.5em{\hfill 0}\hbox to 0.5em{\hfill 0})&\kern-0.7em\hbox to 1.5em{\hfill 3}\hbox to 1.5em{\hfill 7}\hbox to 1.5em{\hfill 11}\hbox to 1.5em{\hfill 9}\hbox to 1.5em{\hfill 7}\hbox to 1.5em{\hfill 5}\hbox to 1.5em{\hfill 3}\hbox to 1.5em{\hfill 2}\hbox to 1.5em{\hfill 1}&0&1155&8&1\\
&(\hbox to 0.5em{\hfill 0}\hbox to 0.5em{\hfill 0}\hbox to 0.5em{\hfill 0}\hbox to 0.5em{\hfill 0}\hbox to 0.5em{\hfill 0}\hbox to 0.5em{\hfill 0}\hbox to 0.5em{\hfill 0}\hbox to 0.5em{\hfill 1}\hbox to 0.5em{\hfill 0})&\kern-0.7em\hbox to 1.5em{\hfill 4}\hbox to 1.5em{\hfill 8}\hbox to 1.5em{\hfill 12}\hbox to 1.5em{\hfill 10}\hbox to 1.5em{\hfill 8}\hbox to 1.5em{\hfill 6}\hbox to 1.5em{\hfill 4}\hbox to 1.5em{\hfill 2}\hbox to 1.5em{\hfill 1}&-2&45&44&1\\
\hline
\end{longtable}

At level one, there is the single three-form corresponding to the root
$r_0$. As there is only a single six-form at level two, the only
sensible level two commutator that can be written down~$(*_1)$
is
\begin{equation}
\label{e2def}
[E_1^{abc},E_1^{def}]=\alpha E_2^{abcdef}.
\end{equation}
As the normalization of $E_2$ can be chosen arbitrarily, we set the
coefficient $\alpha=1$.

The level three, we get a single tensor of mixed symmetry. Again, the
most general ansatz for the level three commutator is
\begin{equation}
\label{level3most-general}
[E_2^{abcdef},E_1^{ghi}]=\left(\alpha_1 E_3^{\begin{Young}a\cr b&c&d&e&f&g&h&i\cr\end{Young}}+\alpha_2 E_3^{\begin{Young}g\cr a&b&c&d&e&f&h&i\cr\end{Young}}\right)\cdot\alpha(abcdef)\cdot\alpha(ghi)
\end{equation}
as there are only two different ways how to distribute indices from
two columns of lengths (6,3) to two columns of lengths (8,1). If we
make use of the identity
\begin{equation}
\begin{array}{ll}
&P\left({\begin{Young}a&b&c&d&e&f&g&h&i\cr\end{Young}}\right)E_3^{\begin{Young}a\cr b&c&d&e&f&g&h&i\cr\end{Young}}=0\\
\Rightarrow&
\left(6 E_3^{\begin{Young}a\cr b&c&d&e&f&g&h&i\cr\end{Young}}
+3 E_3^{\begin{Young}g\cr a&b&c&d&e&f&h&i\cr\end{Young}}\right)\cdot\alpha(abcdef)\cdot\alpha(ghi)=0
\end{array}
\end{equation}
and the freedom in the normalization of $E_3$, this can be reduced to
\begin{equation}
\label{e3def-false-symmetry}
[E_2^{abcdef},E_1^{ghi}]=E_3^{\begin{Young}g\cr a&b&c&d&e&f&h&i\cr\end{Young}}\cdot\alpha(abcdef)\cdot\alpha(ghi).
\end{equation}

In order to obtain an expression for $E_3$ without the extra index
relabeling symmetrizations in terms of commutators $[E_1,E_2]$, we
first apply an index relabeling symmetrization~$(*_2)$
enforcing the column anti-symmetry of 
$E_3^{\begin{Young}g\cr a&b&c&d&e&f&h&i\cr\end{Young}}$:
to the entire expression:
\begin{equation}
\begin{array}{ll}
\label{e3def-resymmetrized}
&[E_2^{abcdef},E_1^{ghi}]\cdot\alpha(g)\cdot\alpha(abcdefhi)\\
=&\frac{1}{3}\left(E_3^{\begin{Young}g\cr a&b&c&d&e&f&h&i\cr\end{Young}}+2E_3^{\begin{Young}a\cr b&c&d&e&f&g&h&i\cr\end{Young}}\right)\cdot\alpha(g)\cdot\alpha(abcdefhi).
\end{array}
\end{equation}
From this point on, we will only work with terms with the proper
column symmetry $\alpha(g)\cdot\alpha(abcdefhi)$.

Besides the desired term, we get an extra contribution where the
tableau structure does not match the index symmetrizations. This can
be eliminated by making use of the Jacobi identity $(*_3)$ in the form
%
\begin{equation}
[[E_1^{abc},E_1^{def}],E_1^{ghi}]+[[E_1^{def},E_1^{ghi}],E_1^{abc}]+[[E_1^{ghi},E_1^{abc}],E_1^{def}]=0.
\end{equation}
which can be reduced to an equation between $E_3$ tensors by use of
the definitions~(\ref{e3def-false-symmetry},\ref{e2def})
\begin{equation}
\begin{array}{l}
\left(E_3^{\begin{Young}a\cr b&c&d&e&f&g&h&i\cr\end{Young}}
-E_3^{\begin{Young}d\cr a&b&c&e&f&g&h&i\cr\end{Young}}
+E_3^{\begin{Young}g\cr a&b&c&d&e&f&h&i\cr\end{Young}}\right)\\
\cdot\alpha(abc)
\cdot\alpha(def)
\cdot\alpha(ghi)=0,
\end{array}
\end{equation}
which then has to be brought to $\alpha(g)\cdot\alpha(abcdefhi)$
symmetry:
\begin{equation}
\label{e3-jacobi}
\left(8E_3^{\begin{Young}a\cr b&c&d&e&f&g&h&i\cr\end{Young}}
+E_3^{\begin{Young}g\cr a&b&c&d&e&f&h&i\cr\end{Young}}
\right)\cdot\alpha(g)\cdot\alpha(abcdefhi)=0
\end{equation}
This can then be used to reduce~(\ref{e3def-resymmetrized}) to
\begin{equation}
\label{e3-def}
[E_2^{abcdef},E_1^{ghi}]\cdot\alpha(g)\cdot\alpha(abcdefhi)=\frac{1}{4}E_3^{\begin{Young}g\cr a&b&c&d&e&f&h&i\cr\end{Young}}
\end{equation}
which is of the desired form.

\subsection{$E_{10}$ at $A_9$ level $\ell=4$}

\noindent When going to level $\ell=4$, a few new features are encountered.
First and foremost, the commutator $[E_3,E_1]$ will now contain two
different representations. The most general ansatz which corresponds
to~(\ref{level3most-general}) for level $\ell=3$ now is
\begin{equation}
\label{level4most-general}
\begin{array}{l}
[E_3^{\begin{Young}a\cr b&c&d&e&f&g&h&i\cr\end{Young}},E_1^{jkl}]\\
=\left(\alpha_1 E_4^{\begin{Young}j&k&l\cr a&b&c&d&e&f&g&h&i\cr\end{Young}}
+\alpha_2 E_4^{\begin{Young}b&j&k\cr a&c&d&e&f&g&h&i&l\cr\end{Young}}\right.\\
\phantom=
+\alpha_3 E_4^{\begin{Young}a&j&k\cr b&c&d&e&f&g&h&i&l\cr\end{Young}}
+\alpha_4 E_4^{\begin{Young}b&c&j\cr a&d&e&f&g&h&i&k&l\cr\end{Young}}\\
\phantom=
+\alpha_5 E_4^{\begin{Young}a&b&j\cr c&d&e&f&g&h&i&k&l\cr\end{Young}}
+\alpha_6 E_4^{\begin{Young}b&c&d\cr a&e&f&g&h&i&j&k&l\cr\end{Young}}\\
\phantom=
+\alpha_7 E_4^{\begin{Young}a&b&c\cr d&e&f&g&h&i&j&k&l\cr\end{Young}}\\
\phantom=
+\beta_1 E_4^{\begin{Young}j\cr k\cr a&b&c&d&e&f&g&h&i&l\cr\end{Young}}
+\beta_2 E_4^{\begin{Young}a\cr j\cr b&c&d&e&f&g&h&i&k&l\cr\end{Young}}\\
\phantom=
+\beta_3 E_4^{\begin{Young}b\cr j\cr a&c&d&e&f&g&h&i&k&l\cr\end{Young}}
+\beta_4 E_4^{\begin{Young}b\cr c\cr a&d&e&f&g&h&i&j&k&l\cr\end{Young}}\\
\phantom=
\left.+\beta_5 E_4^{\begin{Young}a\cr b\cr c&d&e&f&g&h&i&j&k&l\cr\end{Young}}
\right)\cdot\alpha(a)\cdot\alpha(bcdefghil)\cdot\alpha(jkl)
\end{array}
\end{equation}
taking into account all possible ways to re-distribute indices from
the three columns $(a;bcdefghi;jkl)$ to two columns of length $(3,9)$,
resp. three columns of length $(1,1,10)$. Clearly, we have
$\beta_1=0$, due to the anti-symmetry in $jk$. This approach, however,
does not yet fully appreciate the mixed symmetry of the tensors on the
right hand side. Thinking
of~$E_4^{\begin{Young}&&\cr&&&&&&&&\cr\end{Young}}$ as a linear
combination of basis tensors that diagonalize the regular part of the
corresponding Young projector, we can use the fact that sequential
application of any mixed-symmetry projection (i.e. any arbitrary
distribution of the indices to symmetrize and anti-symmetrize over)
corresponding to a tensor of this particular form to any such tensor,
potentially with a different index distribution, followed by Young
projections to the symmetries of the tensors on the left hand side,
yields either zero or (up to an integer multiple that can be absorbed
in the normalization) just one specific combination:
\begin{equation}
\begin{array}{ll}
&54P^{\begin{Young}j&k&l\cr\end{Young}}
 P^{\begin{Young}a\cr b&c&d&e&f&g&h&i\cr\end{Young}}
 P^{\begin{Young}a&j&k\cr b&c&d&e&f&g&h&i&l\cr\end{Young}}
 E_4^{\begin{Young}a&b&c\cr d&e&f&g&h&i&j&k&l\cr\end{Young}}\\
=&
\frac{81}{14}P^{\begin{Young}j&k&l\cr\end{Young}}
 P^{\begin{Young}a\cr b&c&d&e&f&g&h&i\cr\end{Young}}
 P^{\begin{Young}a&j&k\cr b&c&d&e&f&g&h&i&l\cr\end{Young}}
 E_4^{\begin{Young}a&j&k\cr b&c&d&e&f&g&h&i&l\cr\end{Young}}\\
=&\ldots\\
=&
 E_4^{\begin{Young}a&b&c\cr d&e&f&g&h&i&j&k&l\cr\end{Young}}
+2 E_4^{\begin{Young}a&b&j\cr c&d&e&f&g&h&i&k&l\cr\end{Young}}
 E_4^{\begin{Young}a&j&k\cr b&c&d&e&f&g&h&i&l\cr\end{Young}}\\
&
+E_4^{\begin{Young}b&c&d\cr a&e&f&g&h&i&j&k&l\cr\end{Young}}
-2 E_4^{\begin{Young}b&c&j\cr a&d&e&f&g&h&i&k&l\cr\end{Young}}
+E_4^{\begin{Young}b&j&k\cr a&c&d&e&f&g&h&i&l\cr\end{Young}}\\
=:&Z_1\\
\end{array}
\end{equation}

Likewise for the symmetric rank-2 contribution:
\begin{equation}
\begin{array}{ll}
&-108P^{\begin{Young}j&k&l\cr\end{Young}}
 P^{\begin{Young}a\cr b&c&d&e&f&g&h&i\cr\end{Young}}
 P^{\begin{Young}a\cr j\cr b&c&d&e&f&g&h&i&k&l\cr\end{Young}}
 E_4^{\begin{Young}a\cr b\cr c&d&e&f&g&h&i&j&k&l\cr\end{Young}}\\
=&
\frac{81}{2}P^{\begin{Young}j&k&l\cr\end{Young}}
 P^{\begin{Young}a\cr b&c&d&e&f&g&h&i\cr\end{Young}}
 P^{\begin{Young}a\cr j\cr b&c&d&e&f&g&h&i&k&l\cr\end{Young}}
 E_4^{\begin{Young}a\cr j\cr b&c&d&e&f&g&h&i&k&l\cr\end{Young}}\\
=&\ldots\\
=&
-9 E_4^{\begin{Young}a\cr b\cr c&d&e&f&g&h&i&j&k&l\cr\end{Young}}
+9 E_4^{\begin{Young}a\cr j\cr b&c&d&e&f&g&h&i&k&l\cr\end{Young}}\\
&+7 E_4^{\begin{Young}b\cr c\cr a&d&e&f&g&h&i&j&k&l\cr\end{Young}}
+9 E_4^{\begin{Young}b\cr j\cr a&c&d&e&f&g&h&i&k&l\cr\end{Young}}\\
=:&Z_2\\
\end{array}
\end{equation}

Hence, if we start with
\begin{equation}
\label{e4-e3e1}
[E_3^{\begin{Young}a\cr b&c&d&e&f&g&h&i\cr\end{Young}},E_1^{\begin{Young}j&k&l\cr\end{Young}}]=Z_1+Z_2
\end{equation}
we may in principle be able to derive the form of the $[E_2,E_2]$
commutator by making use of the Jacobi identity in the schematic form
$[[E_2,E_1^{(i)}],E_1^{(ii)}]+[[E_1^{(i)},E_1^{(ii)}],E_2]+[[E_1^{(ii)},E_2],E_1^{(i)}]=0$. Before
we discuss this somewhat subtle issue further, we give the result. The
only relevant nonvanishing commutator is:
\begin{equation}
\begin{array}{l}
[E_2^{\begin{Young}a&b&c&d&e&f\cr\end{Young}},
 E_2^{\begin{Young}g&h&i&j&k&l\cr\end{Young}}]\cdot\alpha(abcdefghi)\cdot\alpha(jkl)\\
=
\left(-\frac{1}{28} E_4^{\begin{Young}a&b&c\cr d&e&f&g&h&i&j&k&l\cr\end{Young}}
-\frac{3}{28} E_4^{\begin{Young}a&j&k\cr b&c&d&e&f&g&h&i&l\cr\end{Young}}\right.\\
\phantom=\left.-\frac{3}{28} E_4^{\begin{Young}a&b&j\cr c&d&e&f&g&h&i&k&l\cr\end{Young}}
-\frac{1}{28} E_4^{\begin{Young}j&k&l\cr a&b&c&d&e&f&g&h&i\cr\end{Young}}\right)\\
\cdot\alpha(abcdefghi)\cdot\alpha(jkl)\\
\end{array}
\end{equation}

\subsection{On the role of the Jacobi identity}
\label{l3mod}
\noindent As the structure of the Lie algebra is determined fully
by the Serre relations and the Jacobi identity, and both enter
(indirectly) into the determination of root multiplicities via the
Peterson formula, one may ask the question whether the problem that
the Serre relations and Jacobi identity are interlocked in an unwieldy
way may be circumnavigated by trying to trade the Serre relations for
the known root multiplicities in the level by level determination of
the structure constants. In other words, taking a commutator
definition of the form of~(\ref{e4-e3e1}), what information can be
extracted from the Jacobi identity at the given level? In particular,
is it possible to derive further relations of the
form~(\ref{e3-jacobi}) that can be used to simplify the commutator?

A head-on approach to this problem at level~$\ell$ is to
systematically generate {\em all} possible commutator structures of
$\ell$ tensors $E_1$ and all linear relations between three of them
determined by the Jacobi identity. Taking all possible index
distributions into account, and re-symmetrizing to a given desired
Young tableau column antisymmetry, this is bound to produce all the
relevant relations that can be obtained in such a way. Clearly, this
procedure soon becomes infeasible even on a powerful computer due to
combinatorical explosion, but it is nevertheless important to try this
for the following reasons: First of all, there are many opportunities
where even a small mistake in the implementation of an extensive
algebraic algorithm can lead to wrong results that are hard to detect
with the naked eye, do not occur for many simple test cases, but lead
to inconsistencies in large calculations. An extensive analysis like
the one described is one of the few but crucially important `smoke
tests' available that help to discover program bugs. Unfortunately, it
also can only give binary information: either simplification of a
considerable number of linear equations (typically $\kern0pt>200$
equations containing in total $~1000$ summands with twelve-index
tensors at level four) gives nonsensical relations like a vanishing
commutator that is known to be nonzero, or a set of equations that
seem to make sense. In the former case, the so far only way to find a
bug seems to be to print out a detailed trace of all term
transformations in the program and check this manually -- which easily
becomes highly frustrating sysiphus labour in the long run. The second
reason is that this is the conceptually simplest approach to the
systematic determination of commutators of the form $[E_m,E_{\ell-m}]$
for $m>1$. While one might initially try to start with a definition
for $[E_1,E_{\ell-1}]$ and use the Jacobi identity to derive the
commutator $[E_2,E_{\ell-2}]$ by splitting~$E_{\ell-1}$ to a term of
the schematic form~$\sum [E_1,E_{\ell-2}]$, and then continue with
this procedure to successively determine $[E_3,E_{\ell-3}]$, etc.,
this approach is complicated considerably by the fact that a direct
application requires using a definition of the schematic form
$E_{\ell-m}=\sum[E_1,E_{\ell-m-1}]$, while one starts with an equation
of the form $[E_1,E_{\ell-m-1}]=\sum E_{\ell-m}$, with tensors
belonging to a variety of irreducible representations on the right
hand side, that first has to be solved. Developing the relevant
algorithmic techniques to efficiently automatize this is simplified
greatly by knowing the right answer from the start from a conceptually
simpler and hence more robust approach that may well be not efficient
enough to be used for high levels.

Efficient algorithms to generate a complete set of level~$\ell$
double-commutator identities from the Jacobi identity are described in
appendices~\ref{all-cs},\ref{all-jacobi}. These furthermore have to be
instantiated with all possible distributions of the indices
$abc\,def\,ghi\,\ldots$ to the fundamental $E_1$ tensors, modulo
permutation of groups of three indices,
e.g. $(abc)\leftrightarrow(def)$, and then re-symmetrized in all
possible different ways to the desired symmetry. Special care has to
be taken with substitutions of commutator definitions: it is easy to
overlook that an equation of the form
$[E_\ell,E_m]\cdot\alpha(\ldots)\cdot\alpha(\ldots)=\sum \langle{\rm
tensors}\rangle\cdot\alpha(\ldots)\cdot\alpha(\ldots)$ may only be
substituted into a term containing
$[E_\ell,E_m]\cdot\alpha(\ldots)\cdot\alpha(\ldots)$ after suitable
index re-labeling if the distribution of indices in the
(anti-)symmetrizers $\alpha(\ldots)$ relative to the distribution of
those indices to the columns of $E_\ell, E_m$ can be made to agree.
Efficiently finding suitable index relabelings so that a substitution can be
applied is a further nontrivial task.

Looking at the step from level~$\ell=3$ to~$\ell=4$, the major points
that are not yet visible at level~$\ell=3$ and require generalization
for an algorithmic approach are the following:

$(*_1)$: For a simplified uniform systematic algorithmic treatment,
all commutator definitions have to be given in a form such that both
left hand side and right hand side make the column anti-symmetries of
the tensors in the commutator explicit. Hence, even the $[E_1,E_1]$
commutator should be given in the form
\begin{equation}
\label{e2def}
[E_1^{abc},E_1^{def}]\cdot\alpha(abc)\cdot\alpha(def)=\alpha E_2^{abcdef}\cdot\alpha(abc)\cdot\alpha(def),
\end{equation}
whose anti-symmetrization redundancy might look slightly nonsensical,
but actually helps greatly to simplify many algorithms.

$(*_2)$: One should use a sequence of Young tableau projections
instead of just the corresponding column anti-symmetrizations to
reduce commutator definitions.

$(*_3)$: At least at level~$\ell=4$, trying to use the Jacobi identity
to derive further relations between `wrongly symmetrized' tensors of
mixed symmetry seems to be a red herring. The Jacobi identity does not
produce any new relations of the strict form $\sum\langle{\rm
tensors}\rangle=0$. Even at level~$\ell=3$, it is perhaps more helpful
to think of the corresponding identity as a consequence of a Garnir
symmetry and not the Jacobi identity: for every pair of horizontally
adjacent boxes in a tableau, there is a linear relation between all
tableaus with fillings that can be obtained by permuting the indices
in those two boxes, plus all below the left one, plus all above the
right one, in all possible ways. The coefficients in that linear
relation are $\pm 1$, and determined by the constraint that the
relative sign between terms that differ by an exchange of adjacent
boxes is $-1$. Thus, for the two horizontally adjacent boxes in the
tensor $E_3^{\small\begin{Young}a\cr b&c&d&e&f&g&h&i\cr\end{Young}}$,
we get the relation:
\begin{equation}
\begin{array}{ll}
\phantom+E_3^{\small\begin{Young}a\cr b&c&d&e&f&g&h&i\cr\end{Young}}&
-E_3^{\small\begin{Young}b\cr a&c&d&e&f&g&h&i\cr\end{Young}}\\
-E_3^{\small\begin{Young}c\cr b&a&d&e&f&g&h&i\cr\end{Young}}&
-E_3^{\small\begin{Young}d\cr b&c&a&e&f&g&h&i\cr\end{Young}}\\
-\ldots=0
\end{array}
\end{equation}
Anti-symmetrizing with $\alpha(abcdefhi)$ then
gives~(\ref{e3-jacobi}).  Hence, if the notion of a tensor normal form
already respected Garnir symmetries, the reduction
from~(\ref{e3def-resymmetrized}) to~(\ref{e3-def}) would be automatic.

\subsection{Beyond level four}

\noindent There are a few obvious issues that have to be kept in mind
when going from level four to level five. First of all, one no longer
has one single commutator of the form $[E_4,E_1]$, but two different
level four representations whose commutators with $E_1$ have to be
taken. Furthermore, one no longer can just normalize $E_5$ tensors
such that they appear with coefficient~$1$ in such a defining
equation. The reason is that the $100100000$ representation at
$\ell=5$ may a priori appear in the commutator of both $\ell=4$
representations with $E_1$. One might still argue that it will
presumably at least appear in the commutator of the $200000000$
representation with $E_1$, as this commutator cannot produce the only
other $\ell=5$ representation. If it were zero, the generators
belonging to $200000000$ would have to commute not only with $E_1$,
but (via the Jacobi identity) also with $E_2$, and (inductively) with
all generators at higher level. Hence, the coefficient could be set to
one in this particular commutator. Such reasoning cannot be carried
any further, however, as the $010001000$ representation at level
$\ell=6$ is in the tensor product of both $\ell=5$ representations
with the level-one representation, which both carry further
tensors. At level $\ell=7$, we encounter outer multiplicities $\mu>1$
for the first time.

\section{Conclusion}

\noindent Lacking a powerful theory of the structure of
hyperbolic Kac-Moody algebras, their possible relevance in
gravitational physics -- especially the conjectured role of $E_{10}$
for $M$-theory -- probably justifies going to great lengths to obtain
detailed information about even only the first few elements of a
finite truncation. This is virtually impossible to achieve to
satisfactory depth via manual calculation, both due to the sheer
number of steps and due to the inherent ineptness of humans to execute
them in such large numbers without making mistakes. Hence, this task
is bound to require computer aid. Unfortunately, the algorithmic side
of this problem turns out as being of unusually high complexity, hence
despite considerable technical effort, only very little could be
achieved in this work in terms of new data about $E_{10}$ structure
constants. The most important obstacle to further progress seems to be
the unavailability of powerful tests that simplify the discovery and
analysis of flaws and errors in algorithms and their implementations:
while it is essential to try out various approaches to individual
sub-problems (and, as it turns out, frequently change them when they
have to be embedded into a larger context), all that can be done to
validate an algorithm (not to speak of verification) is to use it to
automatically generate all conceivable relations from a certain not
too small set, semi-automatically check for inconsistencies, and --
should they arise -- eventually redo dozens of lengthy calculations by
hand that involve tensors with more than ten indices, where one has to
especially take great care about signs from index re-orderings. One
has to note that certain problems are specific to situations with
quite many indices, so that simple test cases frequently just do not
exist. The problem may well be that even the appropriate questions
have not yet been found that have to be asked, and answered, to make
the validation phase more bearable. Addressing this problem presumably
should be the most reasonable next step before changing the definition
of normal forms such that Garnir symmetries are properly taken care
of. Then, the issue of a systematic determination of $[E,F]$
commutators has to be addressed before the $E_{10}$ sigma-model can be
extended to higher levels. As it turned out that the task of actually
implementing the relevant algorithms is highly prone to sign and
similar errors, and as debugging is complicated by the fact that it is
virtually impossible to develop an intuition for such terms, a
reference implementation in LISP that also can be used to redo the
calculations presented here within reasonable calculation time will be
made publicly available in the next release
of~\cite{Fischbacher:2002fr}.

\paragraph{Acknowledgments} The analysis presented here could not
have been performed without helpful and encouraging discussions with
Sophie de Buyl, Francois Englert, Marc Henneaux, and Laurent Houart
from ULB, and Hermann Nicolai, and Kasper Peeters, and especially Axel
Kleinschmidt from AEI. The author furthermore acknowledges financial
support by the German Academic Exchange Service via the DAAD postdoc
research program. This work was partially supported by IISN - Belgium
(convention 4.4505.86), by the ``Interuniversity Attraction Poles
Programme -- Belgian Science Policy'' and by the European Commission
FP6 programme MRTN-CT-2004-005104, in which the author is associated
to V.U. Brussel.

\appendix
\renewcommand{\theequation}{\Alph{section}.\arabic{equation}}
\renewcommand{\thesection}{\Alph{section}}

\section{Algorithms}

\noindent In this appendix we list algorithms that are used to address
various parts of the problem.

\subsection{Efficient index (anti-)symmetrization for tensors of high rank}

\noindent The naive translation of the symmetrization/anti-symmetrization
prescriptions for Young tableau projections to a computer algorithm
reaches its limits quite early, often already with problems that --
using some thought -- easily can be done by hand. The primary
underlying reason is that a naive execution of a task like e.g. the
anti-symmetrization over nine indices generates $9!=362\,880$
different individual copies of a term with indices re-distributed in
all possible ways which then usually have to be classified and
subjected to further reductions. Nevertheless, quite many research
problems where such manipulations are necessary are of such a
structure that, given an unlimited amount of time and energy, the
analysis could be taken to almost arbitrary depth. Not seldom, there
is a considerable gap between the point where one can claim to have
obtained a sufficiently deep understanding of the mathematical
structure that going any further is not expected to produce valuable
additional insights and the point where it becomes unfeasible to do
calculations by hand.

This raises the question how to formalize those ideas that make the
simple cases feasible by hand in such a way that they can be executed
as effectively by a machine, only much faster and much more reliable
than by a human. The key idea here is to make as much use of
overlapping symmetrizers and anti-symmetrizers as possible.

\subsubsection{Data Structures}

\noindent While one may want to consider terms formed out of 
general products of tensors with mixed symmetry, we limit our
discussion to the framework of Lie algebras, where the only product of
terms is the commutator. Conceptually, this is also the most
interesting case, as the issue of normal forms of terms will be more
involved here than with simple tensor products. Additionally, we
restrict ourselves to tensors with contravariant indices belonging to
one $SL(n)$ algebra only. This last restriction will eventually have to
be softened, at the very least to allow indices to be either all
contravariant or all covariant, in order to represent commutators of
the $[E,F]$ type. The case of more general base algebras than $SL(n)$
may also be studied, but the price that has to be paid is that the
relation between irreducible representations and corresponding
tableaux gets much more involved, hence this is not considered here.

\paragraph{Tensors} The most natural way to represent a tensor $T$
is as a tuple $T=(\tau, \omega, \gamma, \phi)$, where $\tau$ is the
{\em type} of the tensor, $\omega$ is some abstract weight that will
be used to normalize commutators, $\gamma$ the shape of the Young
tableau, and $\phi$ a vector of indices. On the pragmatic level,
$\tau$ will also include information about a printable tensor name.
We furthermore assume a strict order $\omega<\omega'$ as well as
addition to be defined on tensor weights, which will be used for
sorting tensors, e.g. to determine normal forms for sums.  The tableau
shape $\gamma$ must provide information about where to place an index
from the linearized presentation $\phi$ in the tableau. This can be
achieved e.g. by just recording the lengths of the rows of the
tableau, and using the convention to record indices in normal reading
order. Furthermore, $\gamma$ must provide information that allows one
to quickly extract all the indices belonging to a given column from
$\phi$. When building terms from tensors, it is useful not to put the
tensor $T$ into the term as it is, but rather a tuple $T_*=(T,n)$,
where the tensor is supplemented by an extra bit of information
describing whether reduction to normal form has been applied to this
quantity or not. Note that data representations in a real
implementations may not strictly adhere to these lines, as there might
be practical reasons to e.g. include normalization information
directly in the tensor data structure.

\paragraph{Terms} From tensors, we can recursively form sums and commutators.
This is formalized by the definition of a term~$S$ as a tuple
$S=(s,\Sigma)$, where $s$~is a (sorted) vector of
summands~$(K,\alpha)$ with $K$ either a tensor or a commutator, and
$\alpha$ a coefficient. We impose the restriction that all summands
must be of the same weight. The weight of a commutator is just the sum
of the weights of its summands, and the weight of a term is just the
weight of any of its summands. $\Sigma$ describes all the
non-overlapping symmetrizations or anti-symmetrizations applied to the
indices occurring inside this term. As index-symmetrization is an
operation that is performed by multiplying a
$\delta_{\phantom{(\pm)\;}\ldots}^{(\pm)\;\ldots}$ tensor onto an
expression, this is most naturally represented on the level of terms,
not individual tensors. In particular, symmetrizations will frequently
stretch over indices belonging to two different sides of a commutator.
$\Sigma$ is a list of pairs~$\sigma=(p,v)$ of a type bit discerning
symmetrization from anti-symmetrization and a lexicographically
ordered vector of index names. $\Sigma$ itself is ordered
lexicographically with respect to the index name vectors.

As we may occasionally have to deal with the problem to extract a sign
factor from a term, e.g. when bringing a commutator to normal form,
which would have to be propagated through the `outer context' of a
subterm that is embedded in a term for normalization purposes, it can
be convenient to be able to define normal form modulo a constant
factor, hence we consider augmented terms~$S_*=(S,f,n)$ which are
tuples of a term~$S$ with coefficient of the leading summand
being~$1$, an overall factor~$f$, and a bit denoting whether $S$ and
all its summands are guaranteed to be in normal form.

\paragraph{Commutators} A commutator~$C$ is just a pair $(\ell,r)$ of a left
and a right side, which both are augmented terms $\ell=S_{1*},
r=S_{2*}$.  Again, we augment this structure by an additional bit,
$C_{*}=(C,n)$, that provides information whether both the left and
right entry are guaranteed to be in normal form, and furthermore
whether the left side is smaller than the right side with respect to
the order on terms, to be described next.

\subsubsection{Normal Forms}
\label{normalforms}
\noindent Consequent use of augmented tensors,
commutators, and terms allows to do many symbolic manipulations
without unnecessary normalization steps for subterms that already are
in normal form. Just as the definitions of commutators and terms have
to recursively refer to one another, so do the definitions of their
respective normal forms. We define a term to be in normal form, if its
summands are in normal form and sorted in ascending order with respect
to the order that treats a commutator as smaller than any tensor,
compares commutators lexicographically, and orders tensors first by
their type $\tau$, then by their index distribution (which is assumed
to be normalized), and read for the purpose of ordering in Kanji
reading order (column by column right to left, every column up to
down). A commutator is in normal form if the summands of the left side
are smaller in lexicographical comparison to the summands on the right
side. For index symmetrizations, normal form is defined by first
ordering the indices in every symmetrization block lexically, then
ordering symmetrizations lexicographically by their indices (not
symmetrizing/anti-symmetrizing type).

Commutator and tensor normalization is complicated by the fact that
these may be subject to index symmetrizations that do not match their
own structure (i.e. indices to be antisymmetrized over may be spread
out over both sides of a commutator, or multiple columns of a
tableau). This is resolved as follows: when recursively traversing the
tree of commutators and terms, and mapping it bottom-up to a tree in
normal form, an environmental parameter that denotes the presently
active index symmetrizations is set whenever recursion proceeds from a
term with extra symmetries to its summands. In some programming
languages like Perl or LISP, this is facilitated by the use of local
variables, which, when set in the caller are visible to the callee and
shadowed by inner local definitions. This can prove to be useful in
order to unclutter code from extra function arguments that perhaps
should be kept implicit.

Tensors are normalized by first normalizing all extra index
symmetrizations that act on it to those indices that occur in the
tensor, then applying them sequentially to bring as many lexically
small indices to the rightmost column, then to the second rightmost,
etc. Then, tensor columns are sorted lexicographically, and extra
indices protruding in the first column over the second, which are only
subject to symmetrization from the Young tableau projector, are
brought to inverse lexical order (which will again move the lexically
smallest index to the rightmost column). Clearly, great care has to be
taken with signs introduced by these symmetrizations.

With the normalization of commutators, there are two aspects: first,
which of the terms to place left, and second, how to distribute
indices between them should an index symmetrization from the
environment cover indices on both sides? Unfortunately, these are
somewhat interlocked, and while this issue may be resolved properly,
this would mean both unnecessary code complexity and extra calculation
time for many applications. Hence, we use a somewhat simplistic
approach to this and first compare the weights of the terms on both
sides. Should these turn out to be equal, we do a lexicographical
comparison to the summand vectors, with the same order that is used to
order summands. This is then used to decide which term to put to the
left (the one with larger weight, or the second in lexicographical
comparison for equal weights) in the commutator. Then, we make use of
the presently relevant symmetrizations one after another to rename the
indices in the commutator in such a way that each one tries to bring
as many lexically small indices to the left as possible.

One should keep in mind that with this definition of normal forms,
applying a Young symmetrization to a term containing a single tensor
that is in normal form will in the general case produce a term that
contains more than one tensor in normal form, as in:

\begin{equation}
\begin{array}{l}
P^{\small\begin{Young}j&k&l\cr a&b&c&d&e&f&g&h&i\cr\end{Young}}X^{\small\begin{Young}j&k&l\cr a&b&c&d&e&f&g&h&i\cr\end{Young}}\\
=\left(\frac{7}{10}X^{\small\begin{Young}j&k&l\cr a&b&c&d&e&f&g&h&i\cr\end{Young}}
+\frac{21}{10}X^{\small\begin{Young}a&j&k\cr b&c&d&e&f&g&h&i&l\cr\end{Young}}\right.\\
\left.+\frac{21}{10}X^{\small\begin{Young}a&b&j\cr c&d&e&f&g&h&i&k&l\cr\end{Young}}
+\frac{7}{10}X^{\small\begin{Young}a&b&c\cr d&e&f&g&h&i&j&k&l\cr\end{Young}}\right)\cdot\alpha(abcdefghi)\cdot\alpha(jkl)
\end{array}
\end{equation}

\subsubsection{Symmetrizations}

\noindent The basic operation is to apply an extra index
symmetrization to a term which typically will overlap with some of
those already present. To give a specific example:
\begin{equation}
\begin{array}{l}
X^{\small\begin{Young}a&b&c\cr d&e&f&g&h&i&j&k&l\cr\end{Young}}\cdot\alpha(ajk)\cdot\alpha(bcdefghil)\quad\stackrel{\cdot\alpha(abcdefghi)}\rightarrow\\
\left(-\frac{4}{27}X^{\small\begin{Young}a&j&l\cr b&c&d&e&f&g&h&i&k\cr\end{Young}}
-\frac{14}{27}X^{\small\begin{Young}a&b&j\cr c&d&e&f&g&h&i&k&l\cr\end{Young}}\right.\\
\left.-\frac{2}{27}X^{\small\begin{Young}a&b&l\cr c&d&e&f&g&h&i&j&k\cr\end{Young}}
+\frac{7}{10}X^{\small\begin{Young}a&b&c\cr d&e&f&g&h&i&j&k&l\cr\end{Young}}\right)\cdot\alpha(abcdefghi)\cdot\alpha(jk)
\end{array}
\end{equation}

Taking a term~$S$ that carries~$N$
symmetrizations~$\sigma_j=(p_j,v_j),j\in\{1,\ldots N\}$, where we take
the $v_j$ to cover all of the indices of~$S$ (i.e. we include
anti-symmetrizations over single indices for uniform treatment), it is
comparatively easy to introduce a new symmetrization $\sigma'$ that
completely covers a subset of the existing ones: the new index
symmetries of the re-symmetrized tensor are just the pre-existing ones
plus $\sigma'$ minus those absorbed into $\sigma'$, brought to normal
form. One only has to recursively traverse the term to bring it back
to normal form with the new index relabeling symmetries in place.

For the case that a new symmetrization~$\sigma'$ will only overlap
some of the pre-existing symmetrizations without covering them
completely, a reasonable approach is to first split those overlapping
symmetrizations one by one, and then proceed as above.

Splitting symmetrizers is such an important and ubiquitous function
that a symbolic algebra library should export it to the user. While
one normally would want all such functions to return normalized terms,
it turns out that there are a few situations where such normalization
is not immediately helpful, but computationally expensive, hence it
should be possible to disable the final normalization step with an
appropriate parameter.

Besides the obvious term to be split, the further arguments to the
index splitting function are the previously existing symmetry
$\sigma=(p,v)$ and a (sorted) vector of sub-indices
$v'$. Re-distribution of indices in all possible ways modulo
permutation of those~$n_1$ indices that are in $v'$ and permutation of
those~$n_2$ indices that are in $v$ and not in $v'$ generates -- with
our normalization conventions --
$c=\left(\begin{array}{c}n_1+n_2\\n_1\end{array}\right)$ contributions
obtained by suitable index renamings, with extra coefficients $\pm
1/c$, and -- in the latter case -- if the new index re-distribution is
an even permutation of the former one or not.

Assuming the availability of an iteration construct (to be discussed
later) that for a given pair of natural numbers $(k_1,k_2)$ implements
systematic processing of all different pairs of vectors $(w_1,w_2)$ of
respective lengths $k_1$, $k_2$ for which the concatenation $w_1w_2$
is a permutation of the vector of natural numbers
$[0,\ldots,k_1+k_2-1]$ and for which $w_1,w_2$ are each sorted in
ascending order, index permutation splitting works in detail as
follows:

\begin{tabbing}
\kern2em\=\kern2em\=\kern2em\=\kern2em\=\kern2em\=\kern2em\=\kern2em\\
let\\
\>$h$ be a modifiable (hash) table mapping summands to coefficients,\\
\>$\Sigma'$ be the list of symmetrizations,\\
\>\>with $\sigma$ removed and $(p,v'), (p,v-v')$ added instead\\
\>$p^{a}$ be a vector of~$n_1$ natural numbers\\
\>\> such that $p^{a}_j$ is the position where the $j$-th index of $v'$\\
\>\> appears in $v$,\\
\>$p^{b}$ be a vector of~$n_2$ natural numbers\\
\>\> such that $p^{b}_j$ is the position of the $j$-th index of $v$\\
\>\>that does not appear in $v'$,\\
\>$a$ be the concatenation $p^{a}p^{b}$,\\
\>and~$S'$ be a copy of the term~$S$ with its associated symmetrizations\\
\>\>replaced$^{(*)}$ by $\Sigma'$\\
then\\
\>For all~$c$ ordered choices $(w_1,w_2)$ of vectors of natural numbers\\
\>\>of lengths $\#w_1=n_1$, $\#w_2=n_2$, with $w=w_1w_2$,\\
\>\>\>let\\
\>\>\>\>$v^*$ be the vector with entries $v^*_{a_{w_j}}=v_{a_j}$,\\
\>\>\>\>$\gamma$ be the permutation factor, $\pm1/c$, where the negative sign\\
\>\>\>\>\>only occurs for antisymmetrizations and\\
\>\>\>\>\>when $v^*$ is an odd permutation of $v$,\\
\>\>\>\>$S_{v^*}$ be the term $S'$ where indices are relabeled$^{(**)}$\\
\>\>\>\>\>such that $v_j\mapsto v^*_j$,\\
\>\>\>then\\
\>\>\>\>add all the summands in $S_{v^*}$ multiplied with $\gamma$\\
\>\>\>\>\>(also taking the overall factor $f$ from $S_{v^*}$ into account)\\
\>\>\>\>\>to the result table $h$.\\
then\\
\>Generate a result term from $h$,\\
\>\> and normalize unless expressly wanted otherwise.
\end{tabbing}

Remarks:

$(*)$ {This must happen here since we have to avoid working on terms with inappropriate symmetrizations attached to them in subsequent steps!}

$(**)$ {One can obviously omit performing the relabeling of symmetrizers here.}

\subsection{Generating all bare level $\ell$ commutator structures}
\label{all-cs}
\noindent Given a set of $n$ Lie algebra generators $X_1, X_2,\ldots, X_n$,
we want to generate all recursive commutator structures from them that
are unique up to application of $[A,B]=-[B,A]$. For example, there is
only one such structure at level~$\ell=2$ ($[X_1,X_2]$), while there are
three at~$\ell=3$ ($[[X_1,X_2],X_3], [[X_1,X_3],X_2], [[X_2,X_3],X_1]$)
and~15 at~$\ell=4$ (three of type $[[X_a,X_b],[X_c,X_d]]$ and twelve of
type $[[[X,X],X],X]$).

We inductively define a strict order on commutator structures $P,Q$:

\begin{equation}
P<Q\iff\left\{
\begin{array}{l}
\mbox{$P$ is one of the $X_n$, and $Q$ is $X_m$, and $n<m$, or}\\
\mbox{$P$ is a commutator, and Q is one of the $X_n$, or}\\
\mbox{$P$ and $Q$ are commutators $[P_a,P_b]$ and $[Q_a,Q_b]$}\\
\qquad\mbox{and the vector $(P_a,P_b)$ is lexicographically}\\
\qquad\mbox{smaller than $(Q_a,Q_b)$}
\end{array}
\right.
\end{equation}

We consider a commutator structure to be in normal form if all the
commutator structures contained in it are in normal form, and it is
either one of the $X_n$ or a commutator $[P,Q]$ where $P<Q$.

Given that all commutator structures in normal form of level up to
$\ell=N$ are known, we can form all commutator structures of level up to
$\ell=N+1$ by taking all commutators $[P'_{\ell=j},Q'_{\ell=N+1-j}]$ of
all level-$j$ commutator structures $P'$ with all commutator structures
$Q'$ of complementary level $N+1-j$ smaller than $j$, where $P',Q'$
are formed from $P,Q$ by suitable relabeling of indices. The
$P,Q\rightarrow P',Q'$ index relabeling is done in two steps: first,
one offsets all indices on the $X_a$ in $Q$ by $j$, then one performs
all possible $\scriptsize\left(\begin{array}{c}N+1\\j\end{array}\right)$ index
permutations that preserve the relative order among the indices in the
range $1\ldots j$ and $j+1\ldots N+1$. For $N+1$ even, there are extra
commutators with $P$ and $Q$ of the same level, but one has to take
care to take only those after relabeling where $[P',Q']$ is in normal
form.

The number of different commutator structures grows as follows with
the level:

\begin{equation}
\begin{array}{|r|r|}
\hline
\mbox{Level\hfill}&\mbox{\#Structures\hfill}\\
\hline
1&1\\
2&1\\
3&3\\
4&15\\
5&105\\
6&945\\
7&10\,395\\
8&135\,135\\
9&2\,027\,025\\
\hline
\end{array}
\end{equation}

\subsection{Generating all bare level $\ell$ Jacobi-type relations}
\label{all-jacobi}
\noindent The commutator structures generated by the previous
algorithm are not independent. While they honor the fact that $[A,B]$
and $[B,A]$ are dependent, they do not yet take into account the
Jacobi identity $[[A,B],C]+[[B,C],A]+[[C,A],B]=0$. We want to
systematically generate all such identities between three commutator
structures that can be obtained by applying such a `Jacobi rotation'
to any double commutator. The list of identities generated in this
step may not be algebraically independent, but will be complete and not
contain duplicates.

For an efficient computer algorithm, it is important to make use of
the distinction between `the same' (value in the computer's memory)
and `equal' (with respect to comparison). While it is always easy to
find out if two values are the same, finding out of they are equal may
be costly for recursively built trees, hence the algorithm will
contain canonicalization steps, where a freshly generated commutator
structure is mapped to the one reference value in memory to which it
is equal. We assume our memory representations to be such that two
$X_i$ are equal iff they are the same value, while two commutators
that were generated (hence, memory-allocated) independently at
different steps in the algorithm will be assumed non-equal. 

The input of the algorithm to generate all bare level $\ell$ Jacobi
relations is a list of all level $\ell$ commutator structures, as
generated by the previous algorithm. We in particular assume that no
two commutator structures in this list share any memory
representations for commutators. (Should the implementation of the
previous algorithm not give that guarantee, we first perform a
recursive copy on all structures to make every commutator unique.)

In a first step, we create a dictionary~$C$ that maps (with respect to
being the same, not equality) commutator structures to pairs of an
unique tag (a number) and the corresponding level. This is filled by
recursively walking through all the input commutator structures. This
means that for level~$3$, this table would hold $3\cdot 2$ entries:
the three top-level commutators, and the three inner commutators.

Furthermore, we create a dictionary~$M$ that maps (with respect to
being equal) level $\ell$ commutator structures to their canonical
memory representation, which is just the input data. (For level~$3$,
there would be three entries in this table.) A third dictionary $T$ is
used to remember (with respect to being the equal) pairs of tags of
outer/inner commutators appearing somewhere the canonical memory
representation of commutator structures that already have been used in
some Jacobi relation.

We then process the input commutator structures one after another,
recursively walking through the tree of commutators, where at every
place where we encounter a double commutator of either the form
$[[x,y],z]$ or $[x,[y,z]]$, we obtain the tags of these commutators
from~$C$ and check if these two generators were already used
previously in the generation of a Jacobi relation by checking whether
there already is a corresponding entry in the tag-pair dictionary
$T$. If so, we do nothing at this position and just recurse further
down. If not, we in addition make a corresponding new entry in $T$ and
form the other two sub-structures obtained by cyclically rotating~$x,
y, z$. These are both brought to normal form, taking care of
signs. Note that the subtree depth that is stored in conjunction with
the tag in $C$ is useful to speed up determination of this normal
form.

The problem now is to first form the complete modified level $\ell$
commutator structures for these, and second bring everything to
canonical memory representation. A particularly useful technique to
achieve this that is directly available in functional programming
languages (and has to be emulated in other languages) is {\em
continuation coding}: when recursively traversing the commutator
tree~$X$ down from the root, we hand over two functions as arguments
in the recursive call to the function processing subtrees~$S$: the
first is a buildup function $\beta_S$ that, when called with a a
commutator structure tree $S'$, will generate a commutator structure
that equals $X$ with the subtree $S$ replaced by $S'$. The second is a
selector function $\sigma_S$ that, when called with a commutator
structure tree $\tilde X$ will return (if possible) the subtree
$\tilde S$ of $\tilde X$ that is situated in the same position
relative to the root of $\tilde X$ as is $S$ relative to $X$. In the
recursive call to the left commutator structure subtree $L$ of a
commutator structure $S$, the function arguments $\beta_L, \sigma_L$ are:
\begin{equation}
\begin{array}{lcl}
\beta_L&=&(L'\mapsto \beta_S(\mbox{pair}(L',\mbox{right}(S))))\\
\sigma_L&=&(\tilde X\mapsto\mbox{left}(\sigma_S(\tilde X)))\\
\end{array}
\end{equation}
and likewise for the call to the right commutator structure subtree.

First, the buildup function $\beta_S$ is used twice to generate the
two other Jacobi-rotated level~$\ell$ normalized commutator
structures. These are then mapped to their canonical memory
representation by use of the dictionary~$M$. The selector
function~$\sigma_S$ now allows to locate the canonical memory
representations of the rotated outer and inner commutator. These are
looked up in~$C$ to form their tag pairs, which are then entered
in~$T$ to ensure that the relation that just has been found is not
generated again (twice, even).

The number of relations grows as follows with the level:
\begin{equation}
\begin{array}{|r|r|}
\hline
\mbox{Level\hfill}&\mbox{\#Jacobi relations\hfill}\\
\hline
1&0\\
2&0\\
3&1\\
4&13\\
5&150\\
6&1818\\
7&25\,221\\
8&400\,260\\
\hline
\end{array}
\end{equation}
For level~$\ell=4$, we get explicitly:
\begin{equation}
\begin{array}{lcl}
0=+[[[X_{1},X_{2}],X_{3}],X_{4}]-[[X_{1},X_{2}],[X_{3},X_{4}]]-[[[X_{1},X_{2}],X_{4}],X_{3}]\\
0=+[[[X_{1},X_{4}],X_{2}],X_{3}]-[[X_{1},X_{4}],[X_{2},X_{3}]]-[[[X_{1},X_{4}],X_{3}],X_{2}]\\
0=+[[[X_{1},X_{3}],X_{2}],X_{4}]-[[X_{1},X_{3}],[X_{2},X_{4}]]-[[[X_{1},X_{3}],X_{4}],X_{2}]\\
0=+[[[X_{3},X_{4}],X_{2}],X_{1}]-[[[X_{2},X_{4}],X_{3}],X_{1}]+[[[X_{2},X_{3}],X_{4}],X_{1}]\\
0=+[[[X_{3},X_{4}],X_{1}],X_{2}]-[[[X_{1},X_{4}],X_{3}],X_{2}]+[[[X_{1},X_{3}],X_{4}],X_{2}]\\
0=+[[[X_{2},X_{4}],X_{1}],X_{3}]-[[[X_{1},X_{4}],X_{2}],X_{3}]+[[[X_{1},X_{2}],X_{4}],X_{3}]\\
0=+[[[X_{2},X_{3}],X_{1}],X_{4}]-[[[X_{1},X_{3}],X_{2}],X_{4}]+[[[X_{1},X_{2}],X_{3}],X_{4}]\\
0=+[[X_{1},X_{4}],[X_{2},X_{3}]]+[[[X_{1},X_{4}],X_{3}],X_{2}]-[[[X_{1},X_{4}],X_{2}],X_{3}]\\
0=+[[X_{1},X_{4}],[X_{2},X_{3}]]-[[[X_{2},X_{3}],X_{4}],X_{1}]+[[[X_{2},X_{3}],X_{1}],X_{4}]\\
0=+[[X_{1},X_{3}],[X_{2},X_{4}]]+[[[X_{1},X_{3}],X_{4}],X_{2}]-[[[X_{1},X_{3}],X_{2}],X_{4}]\\
0=+[[X_{1},X_{3}],[X_{2},X_{4}]]-[[[X_{2},X_{4}],X_{3}],X_{1}]+[[[X_{2},X_{4}],X_{1}],X_{3}]\\
0=+[[X_{1},X_{2}],[X_{3},X_{4}]]+[[[X_{1},X_{2}],X_{4}],X_{3}]-[[[X_{1},X_{2}],X_{3}],X_{4}]\\
0=+[[X_{1},X_{2}],[X_{3},X_{4}]]-[[[X_{3},X_{4}],X_{2}],X_{1}]+[[[X_{3},X_{4}],X_{1}],X_{2}]\\
\end{array}
\end{equation}

These equations then have to be instantiated with $X_1=E_1^{abc}$,
$X_2=E_1^{def}$, etc., before definitions of known commutators are
applied to obtain relations between commutators and tensors at the
given level. Then, all different ways how to distribute indices from
the groups $abc$, $def$, etc. to the columns of a given Young tableau
have to be found (modulo exchange of groups). These then have to be
brought to the corresponding tensor's anti-symmetries in index columns.

\end{document}